\documentclass[a4paper]{article}

\usepackage[T1, T2A]{fontenc}
\usepackage[utf8]{inputenc}
\usepackage[english]{babel}

\usepackage[a4paper,top=3cm,bottom=2cm,left=3cm,right=3cm,marginparwidth=1.75cm]{geometry}

\usepackage{amsmath}
\usepackage{amssymb}
\usepackage{graphicx}
\usepackage[colorinlistoftodos]{todonotes}
\usepackage[colorlinks=true, allcolors=blue]{hyperref}
\usepackage{authblk}
\usepackage{amsthm}
\newtheorem{lemm}{Lemma}

\title{Heteroclinic and homoclinic structures in the system of four identical globally coupled phase oscillators with nonpairwise interactions of phases}
\author[1]{\small Evgeny A. Grines}
\author[1]{\small Grigory V. Osipov}
\affil[1]{\footnotesize Lobachevsky   State   University   of   Nizhni   Novgorod,   23
Gagarin av., Nizhny Novgorod 603950, Russia}
\affil[ ]{\textit{evgenij.grines@gmail.com}}
\date{}

\begin{document}

\maketitle

\begin{abstract}
{
Systems of $N$ identical globally coupled phase oscillators can demonstrate a multitude of complex behaviours. Such systems can have chaotic dynamics for $N>4$ when a coupling function is biharmonic. The case $N = 4$ does not possess chaotic attractors when the coupling is biharmonic, but has them when the coupling includes nonpairwise interactions of phases. Previous studies showed that some of chaotic attractors in this system are organized by heteroclinic networks. In present paper we discuss which heteroclinic cycles are forbidden and which are supported by this particular system. We also discuss some of the cases regarding homoclinic trajectories to saddle-foci equilibria.
\\\\}
\end{abstract}

\section*{Introduction}
Systems of interacting elements attract a great interest of specialists in nonlinear dynamics and dynamical systems. The resulting collective behaviour of such systems could be quite complex \cite{PikRos2015,StankPerMcClintStefa2017} even if  a dynamics of a single element is very simple. Studying such systems can often be simplified by approximating them by systems of coupled phase oscillators. 

Systems of identical phase oscillators are a special case of such systems. 
In systems of identical elements the coupling between oscillators is the main source of complexity. 
For example, choosing coupling function that has only the first Fourier harmonic gets a Kuramoto or Kuramoto-Sakaguchi system. 
Although these systems were widely used in studying and explaining phenomenon of synchronization, not all of collective behaviour phenomena can be observed in them; for example, it is not possible to have multiple clusters \cite{EngelMir2014,PikRos2015}.
However, adding a second harmonic already makes dynamics much richer \cite{CluPolRos2016}, allowing cluster solutions \cite{Okuda1993}, attracting heteroclinic cycles \cite{HansMatoMeunier1993,kori2001slow} and chaotic attractors\footnote{This contrasts the case of non-identical oscillators where chaotic attractors can emerge even in the simplest Kuramoto model due to strong detuning of oscillators' natural frequencies \cite{PopMaistrTass2005}.} \cite{AshTownOrosz2007}. 

A case of $N = 4$ phase oscillators is particularly curious. $N=4$ is a minimal number of phase oscillators such that chaotic dynamics might be present in the system. In \cite{AshBickBur2016,AshBickRodr2016} it was stated that there was found no numerical evidences for chaotic behaviour in a system of four identical globally coupled phase oscillators with biharmonic coupling. All known examples with chaotic behaviour for systems of $N=4$ phase oscillators require either higher order Fourier harmonics in coupling function \cite{AshBickTimm2011} or non-pairwise interactions of phases \cite{AshBickRodr2016}. 
It is interesting to understand what makes chaotic examples chaotic in order to find chaos or understand why there is no chaos in case of biharmonic coupling. 

In \cite{AshBickRodr2016} it was noted that for some parameter values attractors of the system (both regular and chaotic) seem to be organized by heteroclinic cycles that consist of equilibria located on invariant hyperplanes in phase space. 
Heteroclinic cycles and homoclinic loops are well-known sources of regular and complex behaviour so it is reasonable to understand what structures are supported or forbidden in the particular system. 
This is a focus of present work. 
The paper is organized as follows. Section \ref{sec:descr-symm} describes a system in consideration and states known facts about its symmetries. 
In Section \ref{sec:little-tech-lemma} we formulate and prove a technical lemma that establishes a relationship between invariant smooth submanifolds of an equilibrium point and invariant linear subspaces of its Jacobi matrix. 
In Section \ref{sec:no-shilnikov-on-face} we prove that there cannot be a homoclinic trajectory to a saddle-focus located on an invariant hyperplane of the system. 
In Section \ref{sec:heteroclinics-on-boundary} we discuss what kind of heteroclinic cycles can be located entirely on invariant hyperplanes of this system and put this in context of attractors that were observed in \cite{AshBickRodr2016}. 
Section \ref{sec:heteroclinic-out-of-boundary} describes heteroclinic cycles that connect equilibria by trajectories not belonging to invariant hyperplanes and discusses how they can influence the dynamics of the system.



\section{The description of a system and its symmetries}
\label{sec:descr-symm}

Consider the following system of $N$ identical globally coupled phase oscillators:
\begin{equation}
\label{eq:full_system}
\begin{split}
\dot{\theta}_j = 
&\tilde{\Omega}(\theta, \varepsilon) + 
\frac{\varepsilon}{N} \sum\limits_{k=1}^{N} g_2 (\theta_k - \theta_j) + \frac{\varepsilon}{N^2}\sum\limits_{k,\,l=1}^{N} g_3(\theta_k + \theta_l - 2 \theta_j) + \\
{} & + \frac{\varepsilon}{N^2}\sum\limits_{k,\,l=1}^{N} g_4(2\theta_k - \theta_l -  \theta_j) + \frac{\varepsilon}{N^3}\sum\limits_{k,\,l,\,m=1}^{N} g_5(\theta_k + \theta_l -  \theta_m - \theta_j),\; j =1, \dots, N,
\end{split}
\end{equation}
where
$$ g_2(\phi) = \xi_1 \cos{(\phi + \chi_1)} + \xi_2 \cos{(2\phi + \chi_2)}, \;\; g_3(\phi) = \xi_3 \cos{(\phi + \chi_3)} , $$
$$ g_4(\phi) = \xi_4 \cos{(\phi + \chi_4)} , \;\; g_5(\phi) = \xi_5 \cos{(\phi + \chi_5)} $$
and $\tilde{\Omega}(\theta, \varepsilon)$ is a symmetric periodic function of phase differences, $\tilde{\Omega}(\theta, 0) = \Omega$. 
The motivation for using such coupling between phase oscillators was given in \cite{AshRodr2014}. 

Recall that a transformation $g\colon \mathbb{R}^n \rightarrow \mathbb{R}^n$ is called a symmetry of a system of ODEs $\dot{x} = F(x)$ if for any solution $\gamma(t)$ of this system $g \circ \gamma(t)$ is also a solution of the system.
The system (\ref{eq:full_system}) has following symmetries \cite{AshSwift1992}:
\begin{itemize}
\item if $(\theta_1(t), \theta_2 (t), \dots, \theta_N (t))$ is a solution, then $(\theta_{\sigma(1)}(t), \theta_{\sigma(2)} (t), \dots, \theta_{\sigma(N)} (t))$ is also a solution
for any permutation $\sigma \in S_N$;
\item if $(\theta_1(t), \theta_2 (t), \dots, \theta_N (t))$ is a solution, then $(\theta_1(t)+ \Theta, \theta_2 (t) + \Theta, \dots, \theta_N (t)+\Theta)$ is also a solution for any $\Theta \in \mathbb{R}$;
\item if $(\theta_1(t), \theta_2 (t), \dots, \theta_N (t))$ is a solution, then $(\theta_1(t)+ 2 \pi k_1, \theta_2 (t) + 2 \pi k_2, \dots, \theta_N (t)+2\pi k_n)$ is also a solution for any $(k_1, k_2 , \dots, k_N) \in \mathbb{Z}^N$. 
\end{itemize}

The ${S}_N$ part of the symmetry group of the full system forces hyperplanes $\lbrace \theta_i = \theta_j \mod 2\pi \rbrace$ to be invariant. Because of that the ordering of phases does not change in time: any permutation $\sigma \in S_N$ and vector $(k_1, k_2, \dots, k_N) \in \mathbb{Z}^N$ correspond to an invariant region defined by inequalities $2\pi k_{1} + \theta_{\sigma(1)} \leqslant 2\pi k_{2} + \theta_{\sigma(2)} \leqslant \dots \leqslant 2\pi k_{N} +\theta_{\sigma(N)} \leqslant \theta_{\sigma(1)} + 2 \pi (k_{1}+1)$. 
However, all such regions can be obtained from each other by applying some transformation from the symmetry group and dynamics in each of these regions is absolutely the same. Thus, we can pick any of these regions as a representative for others.

The $(\theta_1(t), \theta_2 (t), \dots, \theta_N (t)) \mapsto (\theta_1(t)+ \Theta, \theta_2 (t) + \Theta, \dots, \theta_N (t)+\Theta)$ symmetry of system (\ref{eq:full_system}) can be eliminated by passing from phases $\theta_i$ to phase differences $\psi_i = \theta_i - \theta_1$. This leads to a system of equations
\begin{equation}
\label{eq:reduced_system}
\begin{split}
\dot{\psi}_1 &= 0, \\
\dot{\psi}_j &= 
\frac{\varepsilon}{N} \sum\limits_{k=1}^{N} \biggl \lbrack  g_2 (\psi_k - \psi_j) - g_2(\psi_k - \psi_1) \biggr \rbrack 
+ \\
{}& + \frac{\varepsilon}{N^2}\sum\limits_{k,\,l=1}^{N} \biggl \lbrack g_3(\psi_k + \psi_l - 2 \psi_j) - g_3(\psi_k + \psi_l - 2 \psi_1) \biggr \rbrack + \\
{}& + 
\frac{\varepsilon}{N^2}\sum\limits_{k,\,l=1}^{N} \biggl \lbrack g_4(2\psi_k - \psi_l -  \psi_j) - g_4(2\psi_k - \psi_l -  \psi_1) \biggr \rbrack + \\
{}& +
\frac{\varepsilon}{N^3}\sum\limits_{k,\,l,\,m=1}^{N} \biggl \lbrack g_5(\psi_k + \psi_l -  \psi_m - \psi_j) - g_5(\psi_k + \psi_l -  \psi_m - \psi_1)\biggr \rbrack, \; j = 2, \dots, N.
\end{split}
\end{equation}
From $\psi_1 (0) = 0$ and $\dot{\psi}_1(t) \equiv 0$ follows that $\psi_1 (t) \equiv 0$. Thus, the equation for $\psi_1$ can be discarded and phase variables $\psi_2, \psi_3, \dots, \psi_N$ alone describe the behaviour of system (\ref{eq:reduced_system}).

\begin{figure*}
\begin{center}
\includegraphics[height=4.4cm]{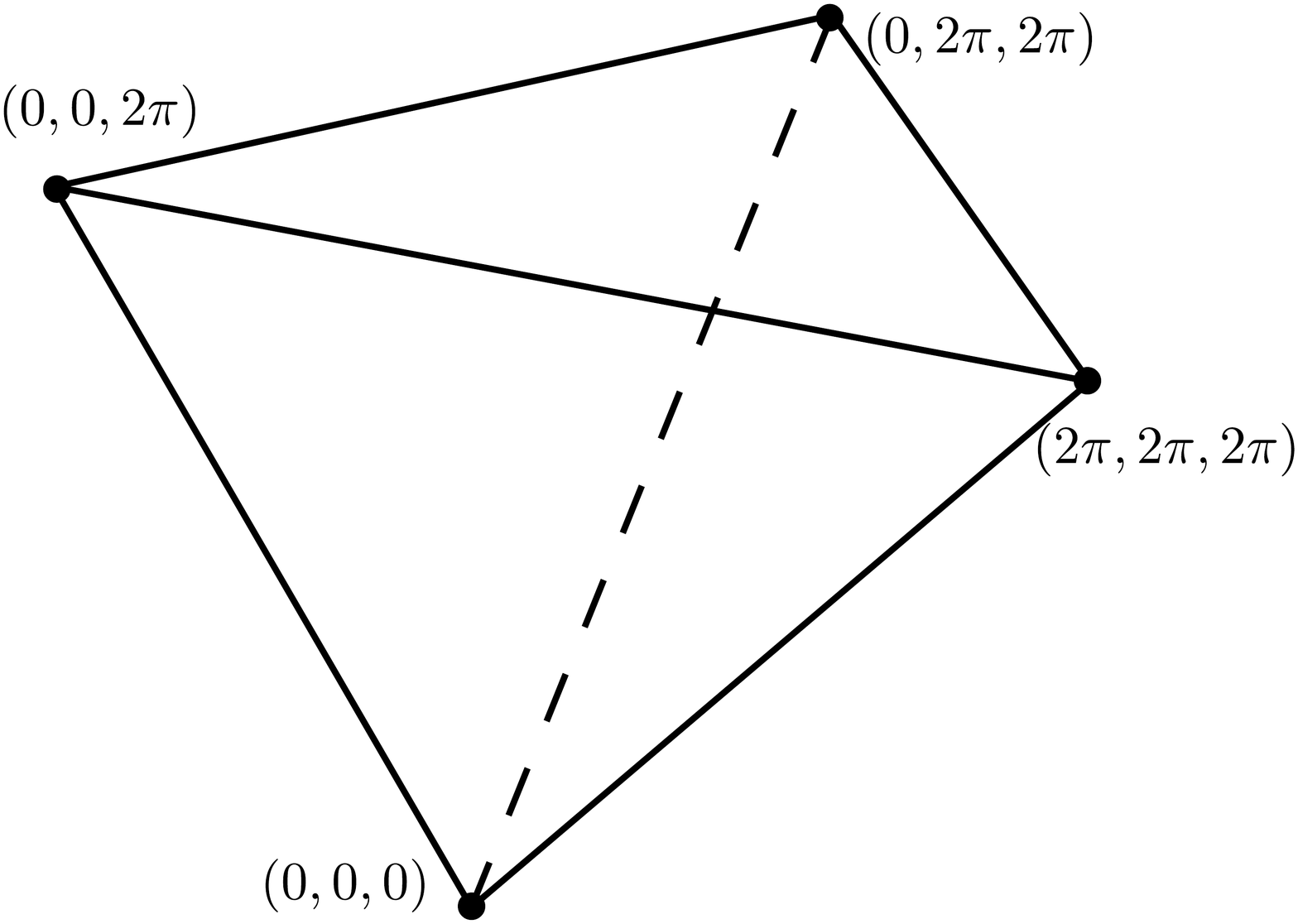}
\end{center}
\caption{The canonical invariant region of system (\ref{eq:reduced_system}).\label{fig:cir}}
\end{figure*}

Let us consider now a case $N=4$ which will be discussed in details in the rest of the paper. Passing to phase differences projects a set $\theta_1 \leqslant \theta_2 \leqslant \theta_3 \leqslant \theta_4 \leqslant \theta_1 + 2\pi$ onto a set $0 = \psi_1 \leqslant \psi_2 \leqslant \psi_3 \leqslant \psi_4 \leqslant 2 \pi$. 
This set is called a \textit{canonical invariant region} $\mathcal{C}$ or CIR \cite{AshSwift1992}. 
Basically, $\mathcal{C}$ is just a tetrahedron in $\mathbb{R}^3$ (see Fig. \ref{fig:cir}), so we can define vertices, edges and faces of a canonical invariant region $\mathcal{C}$ in the same fashion as for tetrahedra.
Note that a reduced system (\ref{eq:reduced_system}) inherits discrete symmetries of system (\ref{eq:full_system}). 
If we keep convention that $\psi_1 = 0$, then a permutation 
$\sigma \in S_4$ and a vector $(k_1, k_2 , k_3) \in \mathbb{Z}^3$
induce a map $$(\psi_2, \psi_3, \psi_4) \mapsto \left (\psi_{\sigma(2)} - \psi_{\sigma(1)} + 2 \pi k_1, \psi_{\sigma(3)} - \psi_{\sigma(1)} + 2 \pi k_2, \psi_{\sigma(4)} - \psi_{\sigma(1)} + 2 \pi k_3 \right ) $$
which is a symmetry of reduced system (\ref{eq:reduced_system}). A canonical invariant region $\mathcal{C}$ has its own symmetry subgroup \cite{AshBickRodr2016,AshSwift1992} generated by a mapping \begin{equation}
\label{eq:CIR-symmetry}
T\colon (\psi_2, \psi_3, \psi_4) \mapsto (\psi_3 - \psi_2, \psi_4 - \psi_2, 2\pi - \psi_2)
\end{equation} and which is isomorphic to $\mathbb{Z}_4$.

\section{Relationship between smooth invariant submanifolds and subspaces of Jacobi matrix of an equilibrium}
\label{sec:little-tech-lemma}
The following lemma
establishes
a connection between known invariant smooth submanifolds that pass through an equilibrium point and invariant subspaces of Jacobi matrix computed at this equilibrium. 

First, let us introduce some notation. Suppose that we have an autonomous system of ODEs $\dot{{x}} = F({x})$, where $F({x})\colon \, \mathbb{R}^n \rightarrow \mathbb{R}^n$ is a sufficiently smooth vector function. 
By $\varphi(t;{x_0})$ we denote a solution of this system that passes through a point ${x_0}$ at $t = 0$, i.e. $\frac{\partial}{\partial t} \varphi(t; {x_0}) \equiv F(\varphi(t; {x_0}))$ and $\varphi(0; {x_0}) = {x_0}$. 
By $DF_{p}$ we denote a Jacobi matrix of a vector function $F(x)$ calculated at point $p$. Now we are ready to formulate the statement of the lemma. 
\begin{lemm}\label{lem:tech-lemma}
Suppose that $\sigma$ is an equilibrium of a system $\dot{x} = F(x)$, $F(\sigma) = 0$. 
Suppose that there is a smooth submanifold $\mathcal{M}$ defined by a system of equations $\{ h_i (x) = 0\}_{i = 1}^{m}, \; h_i \in C^1(\mathbb{R}^n,\, \mathbb{R}),$ and this submanifold passes through the equilibrium $\sigma$, $h_i (\sigma) = 0, \; i=1, \dots, m$. 
Let us also suppose that $\mathcal{M}$ is flow-invariant. 
If all these conditions are met, then $DF_\sigma (T_\sigma \mathcal{M}) = \lbrace DF_{\sigma} v \,\vert\; v \in T_{\sigma}\mathcal{M} \rbrace \subseteq T_\sigma \mathcal{M}$, where $T_\sigma \mathcal{M}$ is a tangent subspace  to the $\mathcal{M}$ at point $\sigma$, $$T_\sigma \mathcal{M} = \{ v \in \mathbb{R}^n \; \vert \; \nabla h_i(\sigma)^T v = 0, \; i = 1, \dots, m \}.$$
\end{lemm}

The proof of this lemma goes as follows.
Pick a tangent vector $v\in T_\sigma \mathcal{M}$.
Since $v$ is a tangent vector to a submanifold $\mathcal{M}$, then there exists a curve $\gamma(s),\; \gamma\colon  (-\delta, +\delta) \rightarrow \mathcal{M}$ such that $\gamma(0) = \sigma$ and $\gamma'(0) = v$. 
Consider a mapping $\Phi(t, s) = \varphi(t; \gamma(s))$ defined for $(t, s) \in (-\varepsilon, +\varepsilon) \times (-\varepsilon, +\varepsilon)$. Since $\mathcal{M}$ is flow-invariant and $\gamma(s) \subset \mathcal{M}$, $\Phi(t, s) \in \mathcal{M}$ for any $(t, s) \in (-\varepsilon, +\varepsilon) \times (-\varepsilon, +\varepsilon)$.
If we fix $t = \tau$, then $\Phi(\tau, s)$ is also a curve that belongs to $\mathcal{M}$ and passes through an equilibrium point, $\Phi(\tau, 0) = \varphi(\tau; \gamma(0)) = \varphi(\tau; \sigma) = \sigma$. Because of that the tangent vector to curve $\Phi(\tau, s)$ at point $\sigma$ lies in $T_\sigma \mathcal{M}$. 
The tangent vector to curve $\Phi(\tau, s)$ at $\sigma$ is simply $\dfrac{\partial \Phi(t, s)}{\partial s} (\tau, 0)$ and since it lies in $T_{\sigma} \mathcal{M}$, then for all $i = 1, \dots, m$ holds $\nabla h_i(\sigma)^T\; \dfrac{\partial \Phi(t, s)}{\partial s} (\tau, 0) = 0 $. 
Moreover, the last equality holds for all $\tau \in (-\varepsilon, +\varepsilon)$, so it's true that $\nabla h_i(\sigma)^T\; \dfrac{\partial \Phi(t, s)}{\partial s} (t, 0) \equiv 0 $. 
Differentiating this by $t$ and substituting $t = 0$ we get that $\nabla h_i(\sigma)^T\; \dfrac{\partial^2 \Phi(t, s)}{\partial t \,\partial s} (0, 0) = 0 $. 
To compute $\dfrac{\partial^2 \Phi(t, s)}{\partial t \,\partial s} (0, 0)$ note that from the definition of $\Phi(t, s)$ immediately follows that $\dfrac{\partial}{\partial t} \Phi(t, s) \equiv F(\Phi(t, s))$. 
Differentiating this expression by $s$ yields 
$\dfrac{\partial^2 \Phi(t, s)}{\partial s \, \partial t} (0, 0) 
= \left . \dfrac{\partial}{\partial s} F(\Phi(0, s))\right \vert_{s = 0} = 
\left . \dfrac{\partial}{\partial s} F(\gamma(s))\right \vert_{s = 0} =
\left . DF_{\gamma(s)} \gamma'(s) \right \vert_{s = 0} = DF_{\sigma}\, \gamma'(0) = DF_{\sigma}\, v$. 
Since we assume that a system of ODEs is smooth enough, partial derivatives commute and we get that for all $i=1, \dots, m$ holds $\nabla h_i(\sigma)^T\; (DF_{\sigma}\, v) = 0$. From this follows that $DF_{\sigma}\, v \in T_{\sigma}\mathcal{M}$. Since $v \in T_{\sigma}\mathcal{M}$ is arbitrary, we get that $DF_\sigma (T_\sigma \mathcal{M}) \subseteq T_\sigma \mathcal{M}$. This concludes the proof of the lemma.

\section{There is no homoclinic trajectories to a saddle-focus on the boundary of CIR}	
\label{sec:no-shilnikov-on-face}

Suppose that $\sigma$ is a saddle-focus that lies on the boundary of CIR. From definition follows that saddle-foci of three-dimensional systems have either two-dimensional stable or unstable manifold. Without loss of generality we can consider that $\sigma$ has a two-dimensional stable manifold.

First, let us prove that a saddle-focus can not be located at edges of CIR. For this we will use 
lemma \ref{lem:tech-lemma}. 
A Jacobi matrix of a saddle-focus equilibrium of three-dimensional system has known structure: it has exactly one invariant one-dimensional subspace and exactly one invariant two-dimensional subspace. 
Let $J_{\sigma}$ denote the Jacobi matrix of system (\ref{eq:reduced_system}) computed at saddle-focus $\sigma$. 
If $\sigma$ was located at a vertex of CIR, then it would have three invariant lines passing through it.
Each line corresponds to its own one-dimensional invariant subspace of $J_{\sigma}$ due to lemma \ref{lem:tech-lemma}, but this contradicts the structure of invariant subspaces of matrix $J_{\sigma}$. 
Now let us suppose that saddle-focus is located at the non-vertex point of edge of CIR. In this case $J_{\sigma}$ must possess two different two-dimensional invariant subspaces due to lemma \ref{lem:tech-lemma}, but that also contradicts structure of invariant subspaces for Jacobi matrix of saddle-focus equilibrium. 
Thus, saddle-foci on the boundary of CIR can be found only in the interior of CIR's faces. 

Let $\Gamma$ denote the face of CIR that contains saddle-focus $\sigma$. Since face $\Gamma$ contains the saddle-focus $\sigma$ and $\Gamma$ is an invariant smooth submanifold, 
due to lemma \ref{lem:tech-lemma} it corresponds to some invariant two-dimensional subspace of matrix $J_{\sigma}$. 
The only two-dimensional subspace of $J_{\sigma}$ corresponds to the stable manifold. 
Since $\Gamma$ is tangent to that subspace, the stable manifold of $\sigma$ must be contained in $\Gamma$ due to Hadamard-Perron's theorem. Thus, the stable manifold of the saddle-focus $\sigma$ lies in face $\Gamma$.

The final step is based on the following observation. 
If there is a homoclinic trajectory to $\sigma$, then the unstable manifold $W^u(\sigma)$ has to intersect the stable manifold $W^s(\sigma)$ at some point $p$. 
Since $p \in W^s(\sigma) \subset \Gamma$ and $\Gamma$ is flow invariant, the trajectory that passes through $p$ stays in $\Gamma$ for all times. 
However, $\sigma$ is a stable focus for the restriction of a system (\ref{eq:reduced_system}) to the face $\Gamma$ and point $p$ cannot tend to $\sigma$ both when $t \rightarrow +\infty$ and $t \rightarrow -\infty$. 
Thus, we have proved that there is no homoclinic trajectories to a saddle-focus on the boundary of CIR\footnote{The similar reasoning proves that a \textit{splay state equilibrium} \cite{AshSwift1992,AshBickRodr2016} of system (\ref{eq:reduced_system}) also cannot have a homoclinic trajectory to itself.}. 



\section{Heteroclinic cycles on the boundary of CIR}
\label{sec:heteroclinics-on-boundary}

\begin{figure}[!ht]
\hspace{1cm}
\begin{minipage}[h]{0.47\linewidth}
\center{\includegraphics[width=1\linewidth]{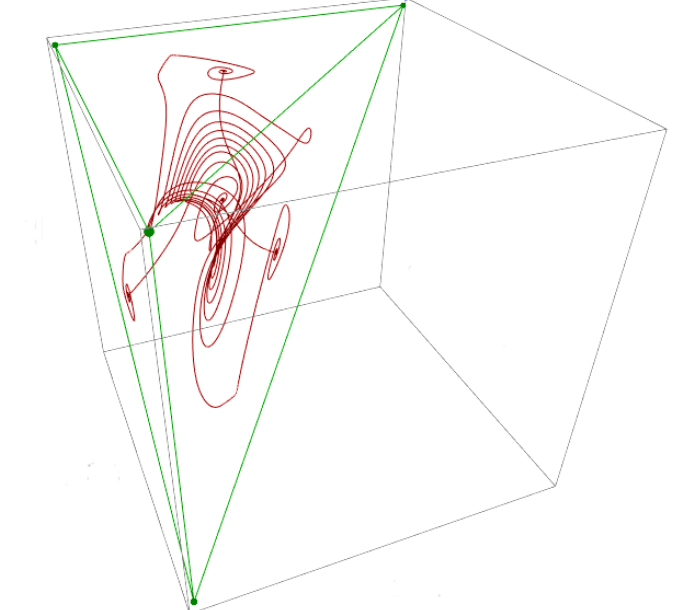} \\ a) }
\end{minipage}
\hfill
\begin{minipage}[h]{0.40\linewidth}
\center{\includegraphics[width=1\linewidth]{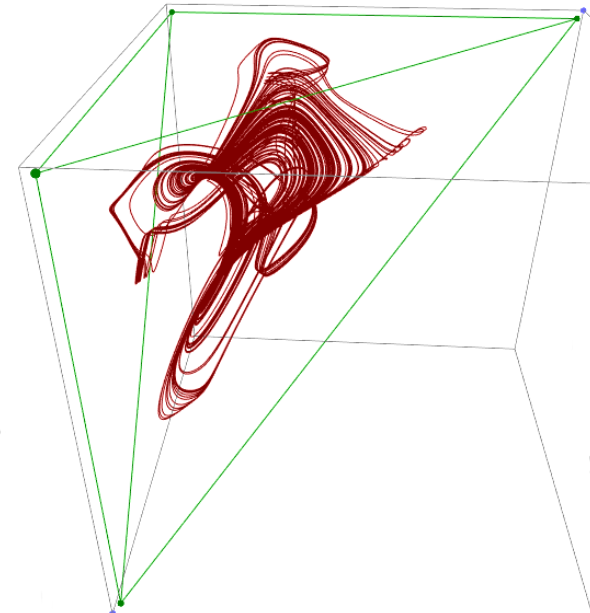} \\ b) }
\end{minipage}
\caption{Attractors of system (\ref{eq:reduced_system}) close to boundary of CIR. a) Limit cycle for $\xi = (-0.3, 0.3, 0.02, 0.8, 0.02), \; \chi = (0.2, 0.316, 0, 1.73, 0)$; b) Chaotic attractor for $\xi = (-0.3, 0.3, 0.02, 0.8, 0.02), \; \chi = (0.154, 0.318, 0, 1.74, 0)$.\label{fig:attractors}}
\end{figure}

In this section we will discuss heteroclinic cycles that lie on the boundary of $\mathcal{C}$. We assume that all equilibrium points are hyperbolic.
While there are plenty of possible options, we focus our attention only on the case when heteroclinic cycle visits all faces of canonical invariant region $\mathcal{C}$. 
The reason behind this is that attractors on Fig. \ref{fig:attractors} are close to all faces of $\mathcal{C}$. 
If such attractors are a result of bifurcation of a heteroclinic cycle, it seems reasonable to think that a heteroclinic cycle itself consists of trajectories and equilibria from all of faces of $\mathcal{C}$.

\begin{figure*}
\begin{center}
\includegraphics[height=6cm]{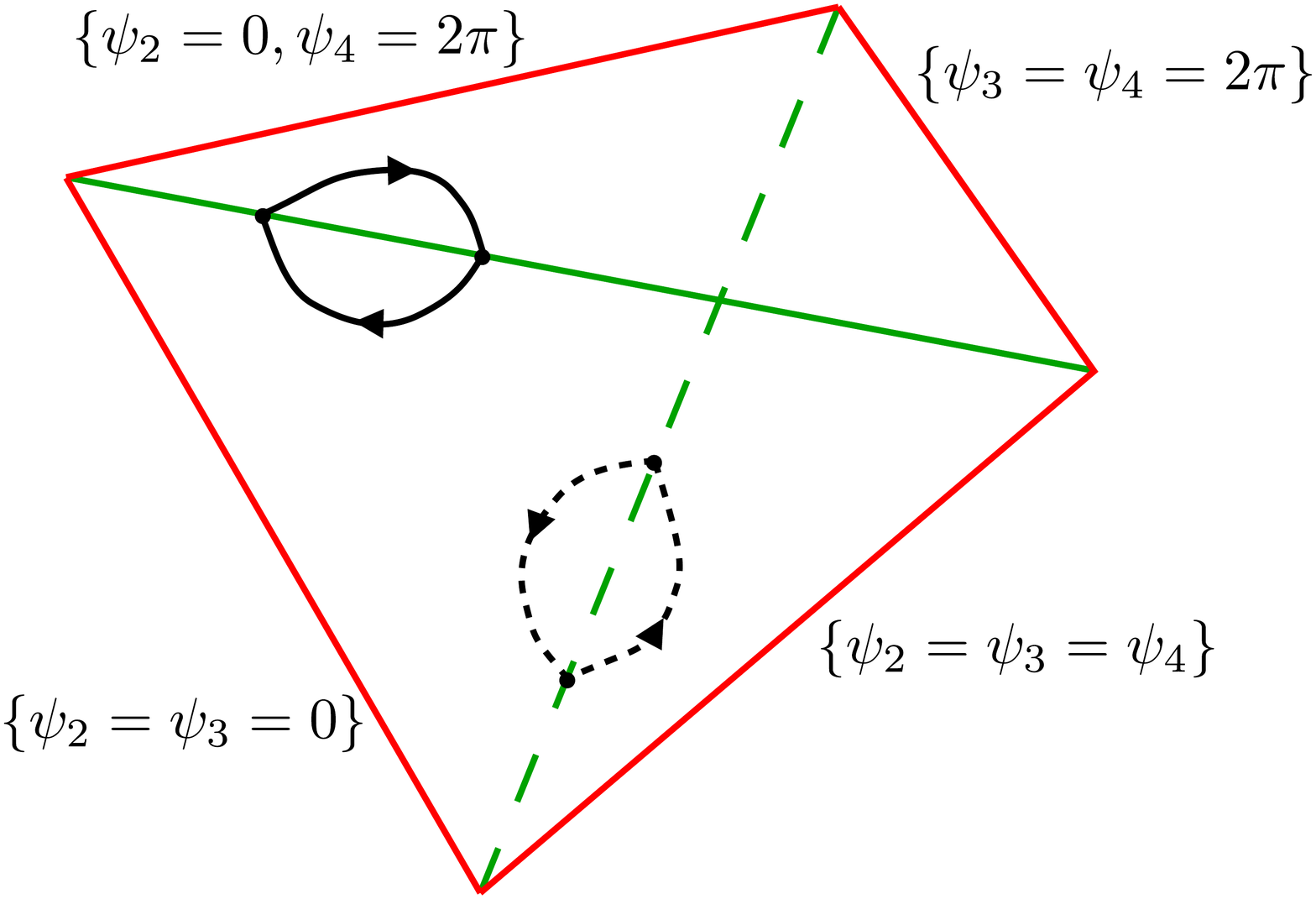}
\end{center}
\caption{If transitions through the red edges of CIR are forbidden, these black-colored trajectories sketch the only possible heteroclinic cycles on the boundary of CIR that involve equilibria at edges. \label{fig:simple_trans}}
\end{figure*}

Note that in order to visit all faces of CIR a heteroclinic cycle on boundary of $\mathcal{C}$ has to intersect at least one of the red edges on Fig. \ref{fig:simple_trans}. Suppose that $\sigma$ is a saddle equilibrium that lies on one of these red edges. 
Without loss of generality we can assume that $\sigma \in  \lbrace \psi_2 = \psi_3 = 0 \rbrace$ since all other red edges can be obtained by applying symmetry (\ref{eq:CIR-symmetry}) to $\lbrace \psi_2 = \psi_3 = 0 \rbrace$.
As a point of CIR $\sigma$ belongs to the intersection of planes $ \{ \psi_2 = 0 \}$ and $ \{ \psi_2=\psi_3  \}$. Since $\sigma$ is a part of heteroclinic cycle, there must be an incoming trajectory $\gamma^{+}$ and an outgoing trajectory $\gamma^{-}$. Without loss of generality we can assume that $W^s(\sigma)$ is one-dimensional and thus $\gamma^{+} \subset W^s(\sigma)$. 
The following mappings (besides being symmetries of reduced system) also form the isotropy group\footnote{If group $G$ acts on a set $X$, an \textit{isotropy group of point $p$} is a subgroup of group $G$ defined by $\lbrace g \in G \, \vert \; g \circ p = p \rbrace$.} of points on line $\lbrace \psi_2 = \psi_3 = 0 \rbrace$:
$$T_0 \colon (\psi_2, \psi_3, \psi_4) \mapsto (\psi_2, \psi_3, \psi_4),$$
$$T_1 \colon (\psi_2, \psi_3, \psi_4) \mapsto (\psi_3, \psi_2, \psi_4),$$
$$T_2 \colon (\psi_2, \psi_3, \psi_4) \mapsto (-\psi_2, \psi_3-\psi_2, \psi_4-\psi_2),$$
$$T_3 \colon (\psi_2, \psi_3, \psi_4) \mapsto (\psi_3-\psi_2, -\psi_2, \psi_4-\psi_2),$$
$$T_4 \colon (\psi_2, \psi_3, \psi_4) \mapsto (-\psi_3, \psi_2-\psi_3, \psi_4-\psi_3),$$
$$T_5 \colon (\psi_2, \psi_3, \psi_4) \mapsto (\psi_2-\psi_3,-\psi_3,  \psi_4-\psi_3).$$
The action of this group on planes $\{ \psi_2 = \psi_3\}$ and $\{ \psi_2  = 0\}$ is summarized in table \ref{table:action}. 
Note that these symmetries are not the symmetries of CIR: they are just symmetries of system (\ref{eq:reduced_system}) in the whole $\mathbb{R}^3$.

\begin{table}[!h]
\centering
\begin{tabular}{|c|c|c|c|c|c|c|}
\hline
~         & $T_0$     & $T_1$        & $T_2$       & $T_3$       & $T_4$       & $T_5$       \\ \hline
$\{\psi_2=0\}$ & $\{\psi_2=0\}$ & $\{\psi_3 = 0\}$  & $\{\psi_2 = 0\}$ & $\{\psi_3 = 0\}$ & $\{\psi_2 = \psi_3\}$ & $\{\psi_2 = \psi_3\}$ \\ \hline
$\{\psi_2=\psi_3\}$ & $\{\psi_2=\psi_3\}$ & $\{\psi_2 = \psi_3\}$  & $\{\psi_3 = 0\}$ & $\{\psi_2 = 0\}$ & $\{\psi_3 = 0\}$ & $\{\psi_2 = 0\}$ \\ \hline
\end{tabular}
\caption{\label{table:action}Action of isotropy subgroup}
\end{table}

\begin{figure}[!ht]
\hspace{1.6cm}
\begin{minipage}[h]{0.30\linewidth}
\center{\includegraphics[height=4cm]{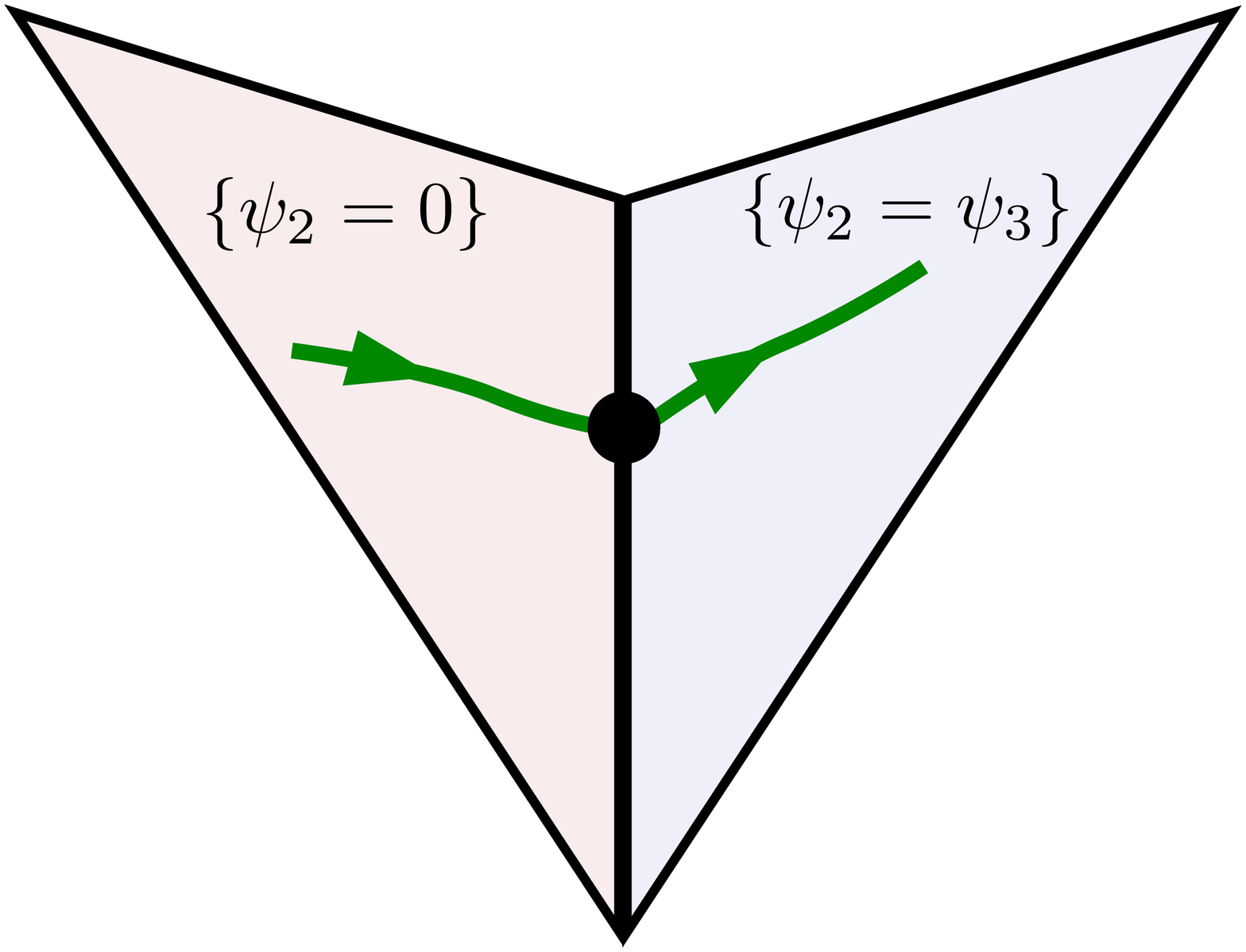}\\ a) }
\end{minipage}
\hfill
\begin{minipage}[h]{0.45\linewidth}
\center{\includegraphics[height=4cm]{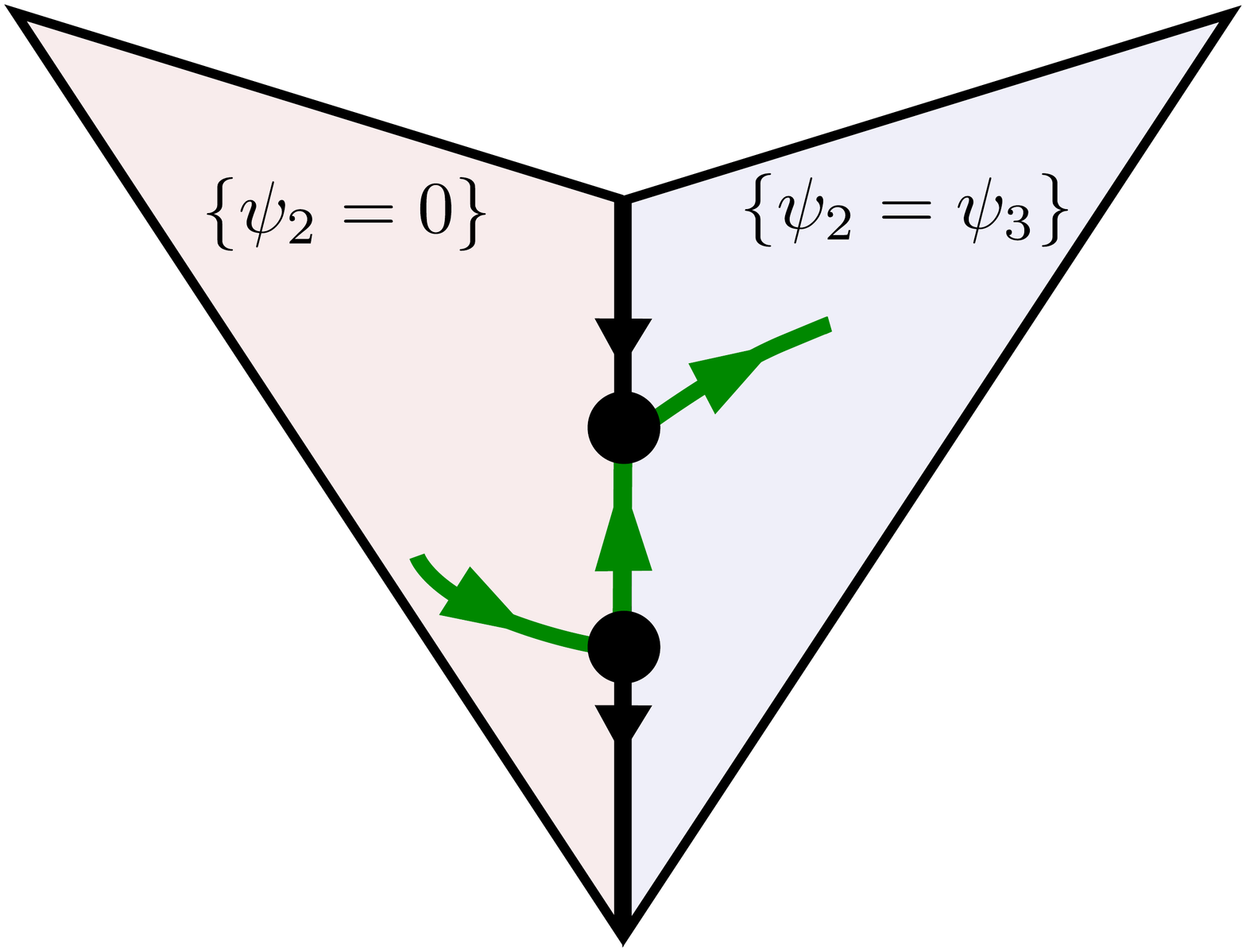} \\ b) }
\end{minipage}
\caption{a) Such transitions from one face of CIR to another are forbidden due to non-trivial isotropy group of points at axis $\lbrace \psi_2 = \psi_3 = 0 \rbrace$; b) Transition from one face of CIR to another that does not contradict symmetry properties. Sketch illustrates a local picture around this pair of equilibria, so other possible equilibria on $\lbrace \psi_2 = \psi_3 = 0 \rbrace$ are not shown.\label{fig:trans}}
\end{figure}

Without loss of generality we can suppose that $W^s(\sigma)$ lies in the plane $\{ \psi_2 = 0\}$. 
In that case there is an element of isotropy group of $\sigma$ that brings a ``copy'' of $W^s(\sigma)$ to a different plane. For example, since $W^s({\sigma}) \subset \{ \psi_2 = 0 \}$, $T_4\circ W^s(\sigma)$ lies in $\{ \psi_3 = 0 \}$. 
However, points from $T_4 \circ W^s(\sigma)$ also tend to $\sigma$ as $t \rightarrow +\infty$ and since $T_4 \circ W^s(\sigma) \cap W^s(\sigma) = \{ \sigma \}$, this violates Hadamard-Perron's theorem: the stable manifold must be unique. 
We arrive at the same conclusion if we suppose that $W^u(\sigma)$ is one-dimensional or one-dimensional (stable or unstable) manifold of $\sigma$ lies in the different plane.
Thus, if $\sigma \in \lbrace \psi_2 = \psi_3 = 0 \rbrace$ is a saddle equilibrium, then its one-dimensional invariant (stable or unstable) manifold must lie in the $\lbrace \psi_2 = \psi_3 = 0 \rbrace$ axis in order to avoid contradiction with symmetry properties. This prevents trajectories to go from one face of CIR to another as pictured at Fig. \ref{fig:trans}a, but option on Fig. \ref{fig:trans}b is not excluded. The same conclusion is true for all other red edges of Fig. \ref{fig:simple_trans}.

As a final remark, let us note that while it is possible to have a heteroclinic cycle with saddle connections like on Fig. \ref{fig:trans}b, such configuration requires both neighbouring equilibria to be saddles. 
As far as we have seen numerically, if we pick any two neighbouring equilibria at edge $\{ \psi_2 = \psi_3 = 0 \}$ one of them is a source or sink. In that case this pair of equilibria cannot be a part of heteroclinic cycle. This probably can be changed if source or sink undergoes some steady-state bifurcation transversely to the edge. So far we don't have any numerical or analytical proof that such heteroclinic cycles are impossible, nor a particular values of parameters at least for which saddle-saddle configuration on an edge can be found. However, if such saddle-saddle configurations are forbidden, the only possible heteroclinic cycles are illustrated on Fig. \ref{fig:simple_trans}. While an example on Fig. \ref{fig:simple_trans} has only two equilibria and two connecting trajectories, such heteroclinic cycles can have more. However, they still can belong only to two faces of $\mathcal{C}$ and we will not consider them in present work. 


\section{Heteroclinic cycles with connecting trajectories in the interior of CIR}
\label{sec:heteroclinic-out-of-boundary}


As we have proven in Section \ref{sec:no-shilnikov-on-face}, if we have a saddle-focus equilibrium on the face of CIR, then its two-dimensional invariant (stable or unstable) manifold necessarily lies in the same face of CIR. 
Heteroclinic cycles that lie on boundary of CIR cannot include such equilibria: the restriction of a system on the face of CIR makes it a source or a sink, hence it lacks either incoming or outgoing trajectory. 
However, such equilibria can be a part of heteroclinic cycle that also has equilibria on all of the faces of CIR, but some of the connecting trajectories lie in the interior of CIR.

\begin{figure}[!ht]
\hspace{1.2cm}
\begin{minipage}[h]{0.30\linewidth}
\center{\includegraphics[height=5cm]{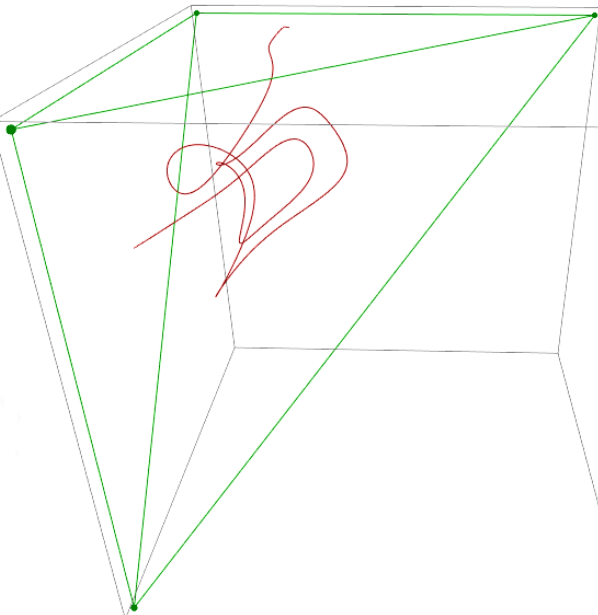} \\ a) }
\end{minipage}
\hfill
\begin{minipage}[h]{0.45\linewidth}
\center{\includegraphics[height=5cm]{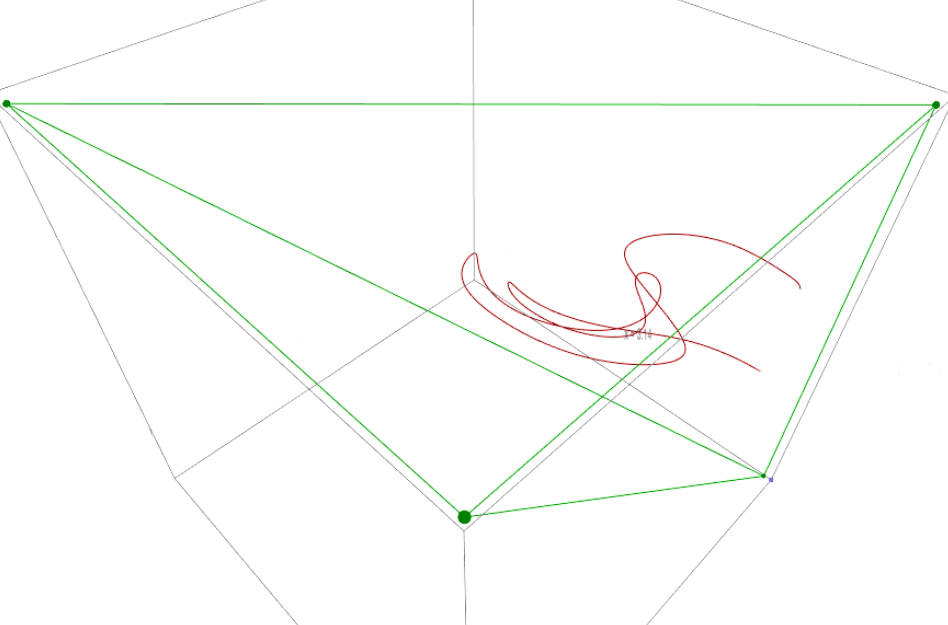} \\ b) }
\end{minipage}
\caption{Heteroclinic trajectories in the interior of $\mathcal{C}$. a) Heteroclinic trajectory connecting a saddle-focus on $\{ \psi_2 = 0 \}$ with a saddle on $\{ \psi_4 = 2\pi \}$; b) Heteroclinic trajectory connecting a saddle-focus and a saddle on the $\{ \psi_2 = 0 \}$.\label{fig:heteroclinics}}
\end{figure}

\begin{figure*}
\begin{center}
\includegraphics[height=8cm]{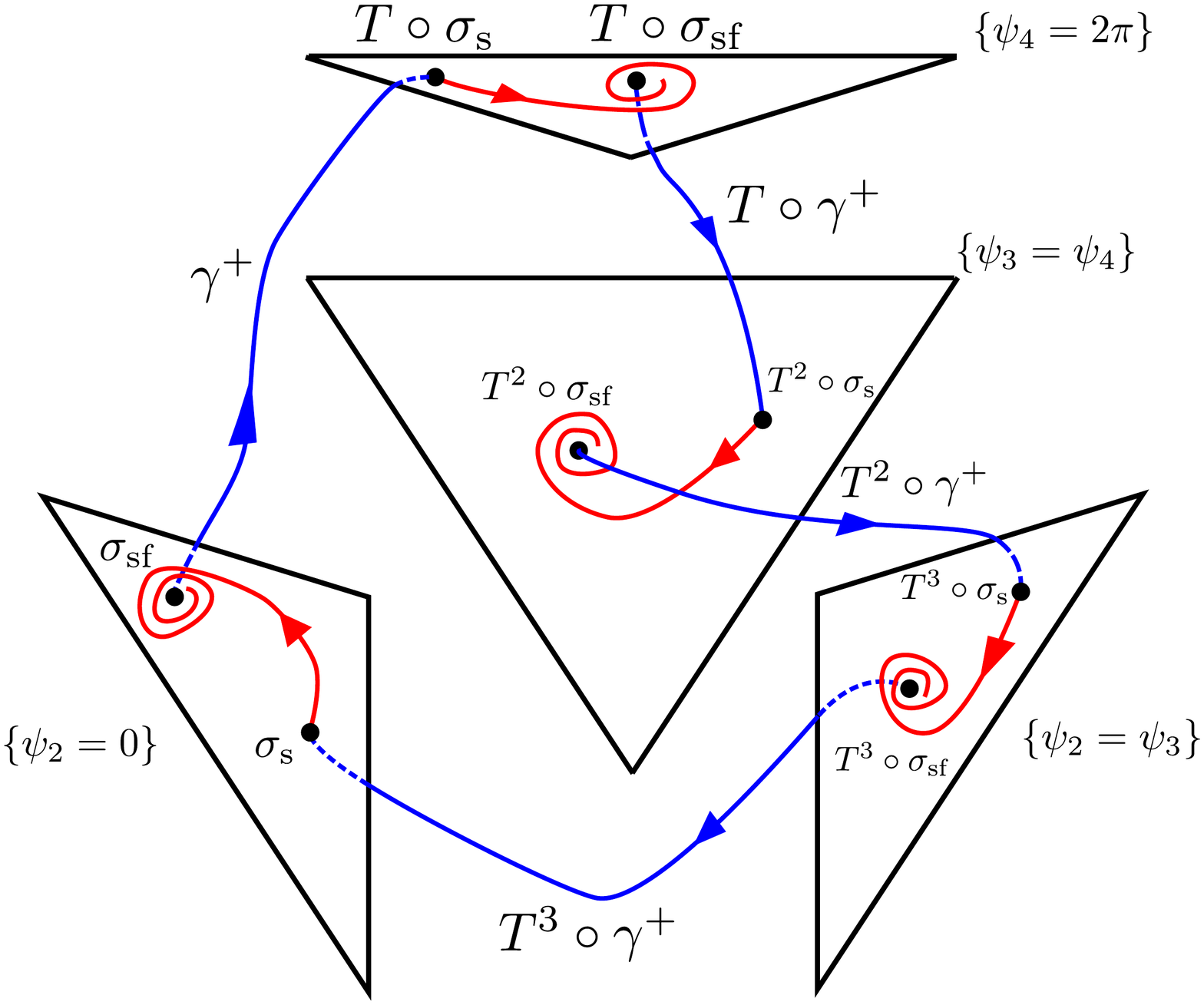}
\end{center}
\caption{A schematic structure of a heteroclinic contour based on Fig. \ref{fig:heteroclinics}a. \label{fig:structure}}
\end{figure*}

Figure \ref{fig:heteroclinics} shows numerically computed examples of such connecting trajectories. In both cases resulting heteroclinic cycle has the following structure (Fig. \ref{fig:structure} illustrates the case a) of Fig. \ref{fig:heteroclinics}). 
A saddle-equilibrium $\sigma_{\rm s}$ lies on a face $\Gamma$ of CIR. For a restriction of dynamics on face $\Gamma$ an equilibrium $\sigma_{\rm s}$ is also a saddle. One of the unstable separatrices of $\sigma_{\rm s}$ goes to a saddle-focus $\sigma_{\rm sf}$ remaining in the face $\Gamma$. An unstable separatrix of saddle-focus $\sigma_{\rm sf}$  (denote it by $\gamma^{+}$) is transversal to a face $\Gamma$ and goes into the interior of CIR. 
Recall that a canonical invariant region has its own symmetry $T$ (\ref{eq:CIR-symmetry}). If $\gamma^{+}$ intersects a stable manifold of $T^{k}(\sigma_{\rm s}), \; k \in \{0, 1, 2, 3 \}$, then due to symmetry generated by mapping $T$ there is a heteroclinic contour that contains equilibria $\sigma_{\rm s}, \, \sigma_{\rm sf},\, T^k (\sigma_{\rm s}),\, T^k(\sigma_{\rm sf}), \dots, T^{3k} (\sigma_{\rm s}),\, T^{3k}(\sigma_{\rm sf}),\; \sigma_{\rm s}$. 
Note that depending on $k$ a heteroclinic contour can consist of 2, 4 and 8 equilibria. For example, case \ref{fig:heteroclinics}a corresponds to $k=1$ or $k=3$, and case \ref{fig:heteroclinics}b corresponds to $k=0$. If $k=2$ a heteroclinic contour will contain 4 equilibria, but we have not observed that numerically yet.

An analytical treatment for such heteroclinic cycles was given in \cite{Tresser1984}. Notice that such heteroclinic cycles are not structurally stable since an intersection of two-dimensional and one-dimensional manifold can be destroyed by a slight change of parameter values. The heteroclinic cycles that we have observed numerically satisfy requirements of Theorem C of \cite{Tresser1984} and thus their destruction leads to an emergence of complex hyperbolic set due to Smale horseshoes for the Poincar\'{e} map near the heteroclinic cycle.






\section{Conclusions}
In present work we have studied what heteroclinic and homoclinic structures are possible in the system of four phase oscillators with nonpairwise coupling. We proved that saddle-foci on faces of CIR can't have homoclinic trajectories. We have proved that symmetries of a system (\ref{eq:reduced_system}) put certain restrictions on connecting trajectories that pass through edges of canonical invariant region and, thus, some types of heteroclinic networks on boundary of CIR are forbidden. It is still an open question whether heteroclinic cycles of type as on Fig. \ref{fig:trans} can be ruled out or not, and how they can influence the dynamics of systems. We also showed that there are heteroclinic cycles whose destruction can lead to an emergence of complex hyperbolic set.

While the system (\ref{eq:full_system}) was a primary topic of this work, we note that the statements we have proved apply not only to this system, but also to all systems with such symmetry group. Probably this can help understanding possibility (or impossibility) of chaotic dynamics in case of biharmonic coupling.



\section{Acknowledgements}
The results of Sections \ref{sec:little-tech-lemma}-\ref{sec:no-shilnikov-on-face} were supported by RSF grant 14-12-00811. The results of Sections \ref{sec:heteroclinics-on-boundary}-\ref{sec:heteroclinic-out-of-boundary} were supported by RFBR grant 16-32-00835.

\bibliographystyle{alpha}
\bibliography{refs}

\end{document}